\documentclass[aps,prd,twocolumn,showpacs,superscriptaddress,groupedaddress]{revtex4}
\usepackage{graphicx}
\usepackage{dcolumn}
\usepackage{bm}
\usepackage{amssymb}
\usepackage{amsmath}
\usepackage{multirow}
\usepackage{mathrsfs}
\usepackage{ulem}
\def\apj{The Astrophysical Journal}

\def\apjs{The Astrophysical Journal Supplement}

\def\prd{Physical Review D}

\def\aap{Astronomy and Astrophysics}

\begin{document}

\title{Limiting Superluminal Neutrino Velocity and Lorentz Invariance Violation by Neutrino Emission from the Blazar TXS 0506+056}

\author{Kai Wang}
 \affiliation{%
 School of Physics, Huazhong University of Science and Technology, Wuhan 430074, China
}%
 \affiliation{%
 Department of Astronomy, School of Physics, Peking University, Beijing 100871, China
}%
 \affiliation{%
 Kavli Institute for Astronomy and Astrophysics, Peking University, Beijing 100871, China
}%
\author{Shao-Qiang Xi}
 \affiliation{%
 School of Astronomy and Space Science, Nanjing University, Nanjing, 210093, China
}%

\author{Lijing Shao}
 \affiliation{%
 Kavli Institute for Astronomy and Astrophysics, Peking University, Beijing 100871, China
}%

\author{Ruo-Yu Liu}%
 \affiliation{%
 School of Astronomy and Space Science, Nanjing University, Nanjing, 210093, China
}%
 \affiliation{%
 Deutsches Elektronen Synchrotron (DESY), Platanenallee 6, D-15738 Zeuthen, Germany
}%
\author{Zhuo Li}
 \affiliation{%
 Department of Astronomy, School of Physics, Peking University, Beijing 100871, China
}%
 \affiliation{%
 Kavli Institute for Astronomy and Astrophysics, Peking University, Beijing 100871, China
}%

\author{Zhong-Kai Zhang}%
 \affiliation{%
 Institute of Geodesy and Geoinformation, University of Bonn, D-53115 Bonn, Germany
}%

\date{\today}

\begin{abstract}
The detection of high-energy neutrino coincident with the blazar TXS 0506+056 provides a unique opportunity to test Lorentz invariance violation (LIV) in the neutrino sector. Thanks to the precisely measured redshift, i.e., $z=0.3365$, the comoving distance of the neutrino source is determined. In this work, we obtain and discuss the constraints on the superluminal neutrino velocity $\delta_\nu$ and the LIV by considering the energy loss of superluminal neutrino during propagation. Given superluminal electron velocity ($\delta_e \ge 0$), a very stringent constraint on superluminal neutrino velocity can be reached,  i.e., $\delta_\nu \lesssim 1.3\times 10^{-18}$, corresponding to the quantum gravity (QG) scale $M_{\rm QG,1}  \gtrsim 5.7 \times 10^{3} M_{\rm Pl}$ and $M_{\rm QG,2}  \gtrsim  9.3 \times 10^{-6} M_{\rm Pl}$ for linear (quadratic) LIV, which are $\sim 12$ orders of magnitude tighter for linear LIV and $\sim 9$ orders tighter for quadratic LIV compared to the time-of-flight constraint from MeV neutrinos of SN 1987A. While given the subluminal electron velocity, a weaker constraint on the superluminal neutrino velocity is obtained, i.e., $\delta_\nu \lesssim 8 \times 10^{-17}$, which is consistent with the conclusions of previous works. We also study the neutrino detection probability due to the distortion of neutrino spectral shape during propagation, which gives slightly weaker constraints than above by a factor of $\sim2$.
\end{abstract}


\maketitle
\section{Introduction}
Lorentz invariance (LI) is the cornerstone of the contemporary theories of fundamental physics, whereas Lorentz invariance may be violated in some candidate theories of quantum gravity (QG) \citep{liberati09, amelino13}. Thus, placing constraints on Lorentz invariance violation (LIV) becomes important to probe the structure of space-time on the Planck scale $M_{\rm Pl}=1.22\times 10^{19}\,\rm GeV$. However, it is huge challenge to test LIV on the Planck scale for the terrestrial experiments \citep{kostelecky11}. For this reason, high-energy astrophysical particles are ideal tools to probe the tiny LIV. QG models which postulate LIV imply a modification in the energy ($E$) -momentum ($p$) dispersion relationship for a particle of rest mass $m$,
\begin{equation}
{E^2} = {p^2} + {m^2} \pm {E^2}{\left( {\frac{E}{{{M_{{\rm QG},n}}}}} \right)^n},
\label{esquare}
\end{equation}
where the $\pm$ sign corresponds to superluminal or subluminal propagation. Regarding the photon sector, the current best limits obtained from the short GRB 090510, according to the arriving time delay among the photons with different energies, are respectively $M_{\rm QG,1} \gtrsim 7.5 M_{\rm Pl}$ and $M_{\rm QG,2} \gtrsim 10^{-8} M_{\rm Pl}$ for linear ($n=1$) and quadratic ($n=2$) LIV \citep{abdo09, vasileiou13}, while Ref.\citep{ellis19} argues such a conclusion drawn from a single GRB may be not robust and based on a systematic study of many sources they conclude the linear LIV is around $0.01-0.1 M_{\rm Pl}$. For the neutrino sector, the generic neutrino LIV operators, at any mass dimension, have been categorized in Ref.~\citep{kostelecky12}. Constraints on linear and quadratic LIV scales are derived as $M_{\rm QG,1} \gtrsim 2.2 \times 10^{-9} M_{\rm Pl}$ and $M_{\rm QG,2} \gtrsim 3.8 \times 10^{-15} M_{\rm Pl} $ for MeV neutrinos of supernova (SN) 1987A \citep{ellis08} and have been considered for the high-energy astrophysical neutrinos observed by the IceCube \citep{diaz14}. Besides, Ref.~\citep{wang16} analyzed the LIV for the possible association (with a relatively low significance) between a PeV neturnio and the gamma-ray flare activity of blazar PKS B1424-418, and set constraints of $M_{\rm QG,1} \gtrsim 0.01 M_{\rm Pl}$ and $M_{\rm QG,2} \gtrsim 6 \times 10^{-8} M_{\rm Pl}$.

In addition to constraints obtained by the time-of-flight delay, in particular, based on an assumed distance of extragalactic neutrino source, Ref.\citep{borriello13} have given constraints on the superluminal neutrino velocity and the LIV for IceCube diffuse neutrinos by treating kinematically allowed energy loss of superluminal neutrino arising from vacuum pair production ($\nu  \to \nu e{e^ + }$, see Section~\ref{sec:loss}), and concluded $\delta_\nu= {v_\nu } - 1 \lesssim \mathcal{O}(10^{-18})$, $M_{\rm QG,1} \gtrsim 10^5 M_{\rm Pl}$ and $M_{\rm QG,2} \gtrsim 10^{-4} M_{\rm Pl}$. Ref.\citep{stecker14} derived $\delta_\nu \lesssim \rm{few}\times10^{-19}$ for IceCube PeV neutrino events and further, Ref.~\citep{stecker15} improved the constraint on superluminal neutrino velocity to $\delta_\nu \lesssim \mathcal{O}(10^{-20})$ by assuming that neutrino sources follow the distribution of star forming rate.

Recently, a track-like neutrino event IceCube-170922A (hereafter, IC-170922A) with energy $\sim 290 \,\rm TeV$ was reported in coincident with a flare of a blazar TXS 0506+056 both spatially and temporally, with a significance at $3\sigma$ level \citep{icecube18a}.
The redshift of blazar TXS 0506+056 has been measured precisely, i.e., $z=0.3365$ \citep{paiano18}, which provides a unique opportunity to constrain the neutrino velocity, as well as the LIV. Some works have used IC-170922A event to constrain the neutrino velocity and the LIV by the time-of-flight delay, e.g., Refs.~\citep{laha18, ellis18, wei18}. In this work, we will examine the constraints on the superluminal neutrino velocity and the corresponding LIV due to the energy loss of vacuum pair production process for IC-170922A event. Our results are summarized in Table~\ref{limits}.

\section{Constraints by IC-170922A}\label{sec:loss}

For the specific case of superluminal neutrinos, three energy loss processes that are otherwise kinematically forbidden , would be allowed even \textsl{in vacuo}, namely, the neutrino Cherenkov radiation ($\nu  \to \nu \gamma$), the neutrino splitting ($\nu  \to \nu \nu \bar \nu $), and the bremsstrahlung of electron-positron pairs ($\nu  \to \nu e{e^ + }$) \citep{cohen11}. The energy of high-energy neutrino will be depleted through these processes during the propagation. Especially, electron-positron pair production is the fastest energy loss process. We can define $\delta_\nu=v_\nu-1$, $\delta_e=v_e-1$ and $\delta_{\nu e}=\delta_\nu-\delta_e$ as in Ref.~\citep{stecker14}, where $c=1$ is the low energy velocity of light \textsl{in vacuo}. For $\delta_\nu \ge \delta_e \ge 0$, the process $\nu  \to \nu e{e^ + }$ is kinematically allowed, which implies ${E_\nu } \ge {m_e}\sqrt {2/{\delta _{\nu e}} }$ \citep{stecker01},  and then the energy loss per unit length determined by this process can be written as ($\hbar  = c = 1$)\citep{cohen11} ,
\begin{equation}
\frac{{dE}}{{dx}} = \frac{{25}}{{56}}\frac{{G_F^2{E^6}{{{\delta^3 _{\nu e} }}}}}{{192{\pi ^3}}} \simeq {\text{1}}{\text{.7}} \times {\text{1}}{{\text{0}}^{{\text{57}}}}{\left( {\frac{E}{{1\,\rm{PeV}}}} \right)^6}\delta _{\nu e}^3\,\rm PeV\,Gp{c^{ - 1}} ,
\end{equation}
where $G_F \simeq 1.2 \times 10^{-5}\,{\rm GeV}^{-2}$ is the Fermi coupling constant. As a result, for a superluminal neutrino with a terminal energy $E_T$, the traveling distance $L$ in the universe has an upper limit, namely, $L \le {\left. {E/(dE/dx)} \right|_{E = {E_T}}}$, so one has
\begin{equation}
{\delta _{\nu e}} \lesssim 8.4 \times {10^{ - 20}}{\left( {\frac{E_T}{{1\,\rm{PeV}}}} \right)^{ - 5/3}}{\left( {\frac{L}{{1\,\rm{Gpc}}}} \right)^{ - 1/3}}.
\label{eq:delta}
\end{equation}

We can obtain the constraint on $\delta_{\nu e}$ from above equation as long as the terminal energy of neutrino and the traveling distance are known. The comoving distance is $D \approx 1.36\,\rm Gpc$ for a redshift $z=0.3365$ by adopting ${H_0} = 67.8\,\rm km/s/Mpc$, ${\Omega _m} = 0.308$, ${\Omega _\Lambda } = 0.692$ \citep{planck16}. For a specific distance of the source, the constraint on $\delta_{\nu e}$ is proportional to the terminal energy of the particle, i.e., $\delta_{\nu e} \propto E^{-5/3}_{T}$.
 Thus accordingly, for the IC-170922A event with a conservative terminal energy $\sim 183 \,\rm TeV$ (the lower limit of energy of IC-170922A reported in \citep{icecube18a}), one has $\delta_{\nu e} \lesssim 1.3 \times 10^{-18}$. So we can obtain the constraint on $\delta_\nu$ once $\delta_e$ is derived. Ref.~\cite{borriello13} concluded a constraint on $\delta_\nu$ based on the assumption $\delta_\nu \gg \delta_e$, while Ref.~\cite{stecker14} considered the possibility that the electron velocity may be superluminal or subluminal. According to constraints on $\delta_e$ given by Ref.~\cite{stecker14} from the Crab nebula, for the constraint on the superluminal electron velocity $0<\delta_e \lesssim 5\times10^{-21}$, we find that $\delta_\nu \simeq \delta_{\nu e}\lesssim 1.3 \times 10^{-18}$ for IC-170922A, and for the constraint on the subluminal electron velocity $-8\times 10^{-17}\lesssim \delta_e <0$ \footnote{A comparable constraint $\mathcal{O}(10^{-17})$ in the electron sector has been given in Ref.~\citep{hohensee13}.}, we find that $\delta_\nu = \delta_{\nu e} +\left| \delta_e \right| \simeq \left| \delta_e \right| \lesssim 8\times 10^{-17}$. Assuming the superluminal electron velocity $\delta_e \ge 0$, the stringent constraint $\delta_\nu \lesssim 1.3\times 10^{-18}$ could be $\sim 9$ orders of magnitude better than the time-of-flight constraint from MeV neutrinos from SN 1987A.

The constraint on LIV can be translated by the constraint on $\delta_\nu$ via the relation \citep{borriello13} \footnote{Here, the definition of $\delta_\nu$ is half of that adopted in Ref.~\citep{borriello13} but same with that used in Ref.~\citep{stecker14}.}
\begin{equation}
{\delta _\nu } \simeq  \pm \frac{1}{2} {\left( {\frac{E}{{{M_{{\rm QG},n}}}}} \right)^n},
\end{equation}
where $\pm$ sign corresponds to the superluminal or subluminal propagation as in Eq.~\ref{esquare}. So, for $\delta_e \ge 0$, we find for the IC-170922A event,
\begin{equation}
M_{\rm QG,1} \gtrsim 5.7 \times 10^{3} M_{\rm Pl},\, M_{\rm QG,2} \gtrsim 9.3 \times 10^{-6} M_{\rm Pl},
\end{equation}
and for $\delta_e \le 0$, we find weaker constraints,
\begin{equation}
M_{\rm QG,1} \gtrsim 94 M_{\rm Pl},\, M_{\rm QG,2} \gtrsim 1.2 \times 10^{-6} M_{\rm Pl}.
\end{equation}
The best constraints in this work on neutrino LIV, i.e., $M_{\rm QG,1}  \gtrsim 5.7 \times 10^{3} M_{\rm Pl}$ and $M_{\rm QG,2}  \gtrsim  9.3 \times 10^{-6} M_{\rm Pl}$, compared to constraints from MeV neutrinos of SN 1987A, are $\sim 12$ orders of magnitude tighter for linear LIV and $\sim 9$ orders tighter for quadratic LIV (The neutronization peak from SN may improve the constraints on LIV for MeV neutrinos of SN, see \footnote{ The constraints of MeV neutrinos of SN could be significantly improved to $M_{\rm QG,1} \sim 10^{-7} M_{\rm Pl}$ and $M_{\rm QG,2} \sim 10^{-14} M_{\rm Pl}$ by the prompt SN $\nu_e$ neutronization burst \cite{kachelriess05} for both super and subluminal neutrino velocity as suggested in Ref.~\citep{chakraborty13}. Such a neutronization peak in the SN neutrino light curve could significantly improve the potential sensitivity of time-of-flight measurements for supernovae (SNe) through the distortion of the observed neutrino's time dispersion, and unlike this case, the vacuum bremsshtrahlung method exploited in this paper is not sensitive at all to the subluminal distortion of the neutrino's dispersion relation but provides much stronger constraints on the superluminal neutrino.}). Note that our constraints on the superluminal neutrino velocity and the LIV are comparable with those given in Refs.~\citep{borriello13}, but an assumed distance of neutrino source was adopted in their works due to the lack of the exact distance information of neutrino emission. Fortunately, the origin of IC-170922A is identified with a correlation to the blazar TXS 0506+056 at a $3 \sigma$ significance level, which allows us to constrain the superluminal neutrino velocity and the LIV more reliably due to the precise measurement of redshift.

However, although neutrinos would lose their energies during propagations, a single neutrino, like IC-170922A, could probably still penetrate through the quantum ``gravity media" and triggers luckily the alert of IceCube, inducing a \textsl{lucky detection}. Due to the probability of lucky detection, the estimation of Eq.~\ref{eq:delta} may be somewhat optimistic because it is based on the typical energy loss length of neutrino equal to the traveling distance. Actually, due to the vacuum bremsstrahlung, the neutrino spectral shape arriving at the Earth should manifest as an exponential cutoff feature. The integrated neutrino event expectation from the neutrino spectrum on Earth can be smaller than 1 but some neutrinos may survive to the Earth. Next, we evaluate the above constraints by considering the neutrino spectral distribution.

For the blazar TXS 0506+056, by adopting the isotropic gamma-ray luminosity between 0.1 and 100 GeV as $ 1.3 \times10^{47}\,\rm erg\, s^{-1}$ and $ 2.8 \times10^{46}\,\rm erg\, s^{-1}$ for $\sim 6$ months period corresponding to the duration of the high-energy gamma-ray flare and the whole observation period of IceCube (i.e., 7.5 years) respectively \citep{icecube18a}, the average integrated gamma-ray fluxes between 0.1 and 100 GeV are $ 3.3 \times10^{-10}\,\rm erg \, {cm}^{-2} s^{-1}$ and $ 7.0 \times10^{-11}\,\rm erg \, {cm}^{-2} s^{-1}$ for two different time periods. Based on the hadronic processes, either photomeson or $pp$ collision, we expect a comparable all-flavor neutrino flux with the gamma-ray flux. The produced gamma-rays with energies larger than TeV will be cascaded to lower energies \citep{liu19}. Besides the neutrino-related gamma-rays, some other relevant processes, e.g., inverse Compton scattering at the source, may contribute additionally the observed gamma-ray flux. High-energy neutrinos can transfer a large fraction of initial energy into $e^{\pm}$ pairs through $\nu  \to \nu e{e^ + }$ and subsequently these high-energy $e^{\pm}$ pairs can convert their energies to the gamma-rays between $\sim$GeV and $\sim$100 GeV through the additional electromagnetic cascades in the cosmic environment \citep{lee98}. As a result, the gamma-ray flux between 0.1 and 100 GeV can be treated as the upper limit of the neutrino flux.

Therefore, the upper limits of the intrinic per-flavor neutrino flux is then $ \sim 1.1 \times10^{-10}\,\rm erg \,{cm}^{-2} s^{-1}$ ($ \sim 2.3 \times10^{-11}\,\rm erg \, {cm}^{-2} s^{-1}$) for an emisssion period of 6 months (7.5 years), after considering the neutrino oscillation. The 90$\%$ confidence level (CL) of the measured energy of IC-170922A is $200 \,\rm TeV$-$7.5 \,\rm PeV$, given a spectral index of -2 \citep{icecube18a}. Invoking the neutrino vacuum bremsstrahlung, in order to derive more conservative constraints, as in Ref.~\citep{borriello13}, we adopt that the expected integrated (anti)muon neutrino detection number can not be smaller than $\sim 0.003$ to guarantee (at $\sim 3 \sigma$) the detection of IC-170922A, so one has
\begin{equation}
t\int_{{E_{\min }}}^{{E_{\max }}} {{A_{eff}}(E)\frac{{d\phi }}{{dE}}} {e^{ - \tau (E)}}dE \gtrsim 0.003,
\label{eq:integ}
\end{equation}
where $t$ is the duration, $A_{eff}(E)$ is the effective area of IceCube \citep{aartsen18} and  $\tau(E)  \simeq {L}/\left( {\frac{E}{{dE/dx}}} \right)$.

Then, we find the constraint on $\delta_{\nu e}$ is only slightly weaker than that given by Eq.~\ref{eq:delta} by a factor of $ \sim 1.8$, inducing slightly weaker constraints on $M_{\rm QG,1}$ by a factor of $ \sim 1.8$ and $M_{\rm QG,2}$ by a factor of $ \sim 1.3$. From Eq.~\ref{eq:delta}, we notice that the energy loss of neutrino is strongly dependent on the neutrino energy, so we tried two other distributions of neutrinos suggested in \citep{icecube18a}, one is with a index of $-2.13$ between $183\,\rm TeV$ and $4.3\,\rm PeV$ and the other is with a index of $-2.5$ between $152\,\rm TeV$ and $2\,\rm PeV$ \footnote{$2\,\rm PeV$ is our assumption, which is not provided by Ref.~\cite{icecube18a} for the index of $-2.5$.}, and found the weaker constraints on $\delta_{\nu e}$ than that given by Eq.~\ref{eq:delta} by a factor of $\sim 2.1$ for former case and a factor of $\sim 2.9$ for latter case. This is because steeper indexes and smaller lower limits of energies will make the neutrino energies concentrate on the lower energy part and lose less energy during propagation. Actually, Eq.~\ref{eq:integ} can be approximately written as $e^{ - \tau} < \sim 0.001$, inducing a weak dependence $\delta_{\nu e} \propto (L/\tau)^{-1/3} $ instead of $\delta_{\nu e} \propto (L)^{-1/3} $ in Eq~\ref{eq:delta}. Since $\tau$ is at most with a value of $\sim$few, which makes the change of constraint small, the constraints obtained by Eq.~\ref{eq:delta} is approximately valid.

\begin{table*}[htpb]
\centering
\caption{Limits on superluminal neutrino velocity and LIV.}\label{limits}
\begin{tabular}{clccc}
\hline\hline

${\delta_e} ^{a}$ & $\delta_{\nu} $ & $M_{\mathrm{QG},1}(M_{\rm Pl})$ & $M_{\mathrm{QG},2}(M_{\rm Pl}) $ \\
\hline
$0 \leq \delta_e \lesssim 5\times10^{-21}$ & $1.3\times 10^{-18}$ & $5.7 \times 10^{3}$ & $9.3 \times 10^{-6}$ \\
\hline
$-8\times 10^{-17}\lesssim \delta_e <0$ & $8\times 10^{-17}$ & $94$ & $1.2 \times 10^{-6}$ \\
\hline

\end{tabular}
\begin{list}{}
\centering
\item{$a$}: $\delta_e \lesssim 5\times10^{-21}$ for $\delta_e \ge 0 $ and $\left| \delta_e \right|  \lesssim 8\times10^{-17}$ for $\delta_e  \le  0 $ are from the constraints given in Ref~\citep{stecker14} based on the observation of Crab nebula.\\
\end{list}
\end{table*}

\section{Conclusions}
The detection of high-energy neutrino event IC-170922A in coincidence with the blazar TXS 0506+056 is the first time in history to identify the direct correlation between the high-energy neutrino and the astrophysical source at a significance level of $3\sigma$. Such a correlation allows us to probe the fundamental physics, e.g., the neutrino velocity and LIV. The resulting constraints on the superluminal neutrino velocity and the LIV are summarized in Table~\ref{limits}.

For the usual method to constrain the LIV by the time-of-flight delay, the exact correlation between gamma-rays and neutrinos is required, therefore the obtained results are limited by the uncertainty of such a correlation. To avoid these uncertainties, in Section~\ref{sec:loss}, we adopt a direct method, i.e., considering neutrino energy loss during propagation. For this method, the constraint on superluminal neutrino velocity in this work can reach a level of $\delta_\nu \lesssim 1.3\times 10^{-18}$ by assuming the superluminal electron velocity $\delta_e \ge 0$.
The corresponding QG scales in the neutrino sector $M_{\rm QG,1}  \gtrsim 5.7 \times 10^{3} M_{\rm Pl}$ and $M_{\rm QG,2}  \gtrsim  9.3 \times 10^{-6} M_{\rm Pl}$, are $\sim 12$ orders of magnitude tighter for linear LIV and $\sim 9$ orders tighter for quadratic LIV compared to the time-of-flight constraint from MeV neutrinos of SN 1987A.
In addition, for the subluminal electron velocity $\delta_e \le 0$, the constraint on superluminal neutrino velocity is determined by the limit of electron velocity, i.e., $\delta_\nu \simeq \left| \delta_e \right|$, which gives a similar conclusion as in Ref.~\citep{stecker14}. Taking the possible lucky detection into account, we have calculated the integrated neutrino detection number by considering the distortion of neutrino spectrum due to the vacuum bremsstrahlung. We set a criterion that integrated neutrino detection number should be larger than $0.003$ to guarantee (at $\sim 3\sigma$) the detection of IC-170922A and obtain the constraints which are slightly weaker than above constraints by a factor of $\sim 2$.

For the neutrino source with a specific distance, the constraint on the LIV is proportional to the energy of neutrino and therefore if in the future the higher energy neutrino is detected, a more stringent limit can be expected. For the future EeV ($10^{18}\,\rm eV$) neutrino experiments, e.g., ARA and ARIANNA \citep{barwick15, ishihara15}, they have abilities to capture very-high-energy cosmogenic neutrinos \citep{wang17}, which could improve constraints on the superluminal neutrino velocity and the LIV significantly.

\begin{acknowledgments}
We thank the anonymous referee for constructive comments that have allowed us to express our intentions more appropriately. This work is supported by the NSFC grant 11773003, 973 program grant 2014CB845800, the China Postdoctoral Science Foundation (No. 2019M650311) and the Fundamental Research Funds for the Central Universities (No. 2020kfyXJJS039).
\end{acknowledgments}

\end{document}